# r/K selection of GC content in prokaryotes

Lucio Aliperti Car[1], Gonzalo Farfañuk[1], Luciana L. Couso[2], Alfonso Soler-Bistué[3], Ariel A. Aptekmann[4], Ignacio E. Sánchez[1]

[1] Universidad de Buenos Aires. Consejo Nacional de Investigaciones Científicas y Técnicas. Instituto de Química Biológica de la Facultad de Ciencias Exactas y Naturales (IQUIBICEN). Facultad de Ciencias Exactas y Naturales. Laboratorio de Fisiología de Proteínas. Buenos Aires, Argentina.
[2] Universidad de Buenos Aires. Facultad de Agronomía. Cátedra de Genética. Buenos Aires, Argentina
[3] Instituto de Investigaciones Biotecnológicas Dr. Rodolfo A. Ugalde, CONICET, Universidad Nacional de San Martín, Argentina.
[4] Marine and Coastal Sciences Department, Rutgers University
Correspondence can be addressed to I.E.S.: isanchez@qb.fcen.uba.ar

**ABSTRACT**

The GC content of prokaryotic genomes is species-specific and takes values from 16 to 77 percent. There are currently no accepted explanations for this diversity of selection for GC content. We analyzed the known correlations between GC content, genome size and amino acid cost in thousands of prokaryotes, together with new or recently compiled data on cell shape and volume, duplication times, motility, nutrient assimilation, sporulation, defense mechanisms and Gram staining. GC content integrates well with these traits into r/K selection theory when phenotypic plasticity is considered. High GC content prokaryotes are r-strategists with cheaper descendants thanks to a lower average amino acid metabolic cost and a smaller cell volume, colonize unstable environments thanks to flagella and a bacillus form and are generalists in terms of resource opportunism and the ability to defend themselves from various hazards. Low GC content prokaryotes are K-strategists specialized for stable environments that maintain homeostasis via a high-cost outer cell membrane and endospore formation as a response to nutrient deprivation and attain a higher nutrient-to-biomass yield. The lower proteome cost of high GC content prokaryotes is driven by the association between GC-rich codons and cheaper amino acids in the genetic code, while the correlation between GC content and genome size may be partly due to a shift in the functional repertoire of genomes driven by r/K selection. In all, we show that molecular diversity in the GC content of prokaryotes and the corresponding species diversity may be a consequence of ecological r/K selection.

**INTRODUCTION**

The GC content of a genome is the fraction of bases that are either guanine (G) or cytosine (C). It is a stable trait of a given prokaryotic genome, taking values from 16 to 77% (1). Molecular mechanisms leading to both higher (2) and lower (3) GC contents have been described. However, mutation alone cannot explain the observed base compositions, implying a role for natural selection in determining genomic GC content (4). It has been proposed that higher GC content correlates with higher growth temperatures due to a higher stability of the DNA double helix. However, this hypothesis did not hold true in a general manner (5). Alternatively, variations in GC content of genomes and the corresponding transcriptomes may result from the interplay between nutrient availability and nucleotide metabolic cost. The synthesis of genomes and transcriptomes that are rich in adenine (A) and thymine/uracil (T/U) would have a lower metabolic cost according to some reports (6). In all, selection for a particular value of GC content in each organism remains largely enigmatic. Several traits of prokaryotes are known to vary consistently with genomic GC content. GC content correlates strongly with codon usage, so that GC-rich codons are more common in GC-rich genomes and vice versa (7). In turn, this leads to differences in amino acid usage between proteomes coded by AT-rich and GC-rich genomes (7). Higher GC content has been linked to larger genomes (8), whose gene repertoire may increase the ability to grow in unstable environments (9) through metabolic and sensing versatility (10–12) and the ability to cope with hazards (13). However, the relationship between GC content, metabolic versatility and defense mechanisms has not been explored in depth.

In ecology, certain combinations of traits such as GC content, genome size and proteome metabolic cost are usually considered in terms of strategies (10, 14). A classic framework is r/K selection theory (15, 16), which was devised to understand island biodiversity and later applied to other cases of density-dependent natural selection, including prokaryotes (17). Examination of the logistic population equation helped define two archetypal organisms in terms of the maximum growth rate r (higher in r-strategists) and the carrying capacity K of the local environment (higher in K-strategists). Empirical studies have found additional traits commonly associated to these archetypes. For example, K-strategists often produce offspring of higher quality and cost and at a lower rate, while r-strategists often produce cheaper, lower quality descendants at a faster rate (15, 16). In the case of prokaryotes, r-strategists are expected to present a high migratory tendency (17), a low yield of nutrient conversion to biomass (17, 18) and a low specific affinity for substrates (17, 18). The relationship between the number of genes related to specific functional categories and r/K selection has only been tested in a few specific environments and so far did not yield itself to a general interpretation (19, 20). To our knowledge, genomic GC content in prokaryotes has not been considered in depth as a potential trait in r/K selection.

r/K selection can be understood as a trade-off (21) between rapid growth in unstable environments and high population in stable environments (22). In this view, K-strategists are also termed gleaners and r-strategists are also termed opportunists (22). Phenotypic plasticity is the ability of a single genotype to produce different phenotypes upon variations in the environment (23) and, as any other phenotypic trait, can be subject to selection (24–26). Selection in unstable environments may favor plastic phenotypes, while selection in predictable habitats may favor fixed phenotypes (27). Similarly, phenotypic plasticity is expected to underlie greater niche breadth in generalist organisms, while specialists might be less plastic (28). We hypothesized that prokaryotic r-strategists not only produce cheaper descendants with a low yield but also repeatedly colonize unstable environments through spatial mobility and are generalists with a high degree of proteome phenotypic plasticity. This proteome plasticity likely enables the use the different resources that successively become available in an unstable environment, i.e., resource opportunism (22, 29), and the ability to defend themselves from the various hazards that successively present themselves in an unstable environment (30). On the other hand, we expected prokaryotic K-strategists not only to produce more expensive descendants with a high yield but also to be specialists for individual stable environments with a lower number of available resources and a relatively short list of environmental hazards. In such a situation, a high fitness may be more easily attained with low levels of motility and phenotypic plasticity. Prokaryotic r- and K-strategists would balance each other depending on the cost versus opportunity balance of energy investments into offspring size and cost, motility, resource opportunism and defense mechanisms in stable versus unstable environments.

Our global analysis of phenotypic traits that vary with GC content took into account previously reported and newly generated data (see methods) for reproductive traits (cell volume (31), proteome cost (32), doubling times, growth temperature), motility traits (cell shape (31), presence of motility and flagella (31)), genome size (33), genes associated to nutrient sensing, uptake and assimilation (12, 34) and endospore formation (31), yield for conversion of nutrients to biomass (35, 36), genes associated to defense functions (12, 34) and Gram staining (31). We found that r/K selection theory and phenotypic plasticity can rationalize variations in genomic GC content in prokaryotes.

## RESULTS

### GC content in prokaryotes and associated traits

We have compiled the GC content of coding sequences for over 49000 bacteria and 700 archaea (see methods and Supplementary File 1). Coding sequence GC content closely follows genomic GC content (Supplementary Figure 4, Spearman's rank correlation coefficient r is 0.99, $p^* < 9 \cdot 10^{-200}$). GC content in prokaryotes varies from 16 to 77%, with 90% of organisms presenting a GC content from 33 to 71% (Figure 1A). These empirical limits are narrower in eukaryotic genomes, where GC content takes values from 21 to 71%, with 90% of organisms presenting a GC content from 37 to 59% (Figure 1A and Supplementary File 1). In our database, high GC content correlates with lower growth temperatures for 6695 prokaryotes (Figure 1B, r -0.19, $p^*$-value $5 \cdot 10^{-57}$, p-values corrected for multiple testing as described in methods), in agreement with previous results (5). We also examined the metabolic cost of AT(AU) relative to that of GC as reported in five different sources (37–41). The relationship between metabolic cost of AT(AU) and GC content depends on the details of the calculation (Supplementary File 2), suggesting that genomes and transcriptomes with a high GC content do not necessarily present a higher metabolic cost. Last, we used amino acid costs from our previous work (32) to confirm previous reports (37) that GC-rich codons code for amino acids with a lower metabolic cost (Figure 1C and Supplementary File 3, r -0.37, $p^*$ $4 \cdot 10^{-3}$). This suggests that the synthesis cost of proteomes from GC-rich genomes could be smaller. However, the bimodality in the amino acid cost for codons with 1 and 2 GC, shows that this depends on the details of amino acid usage.

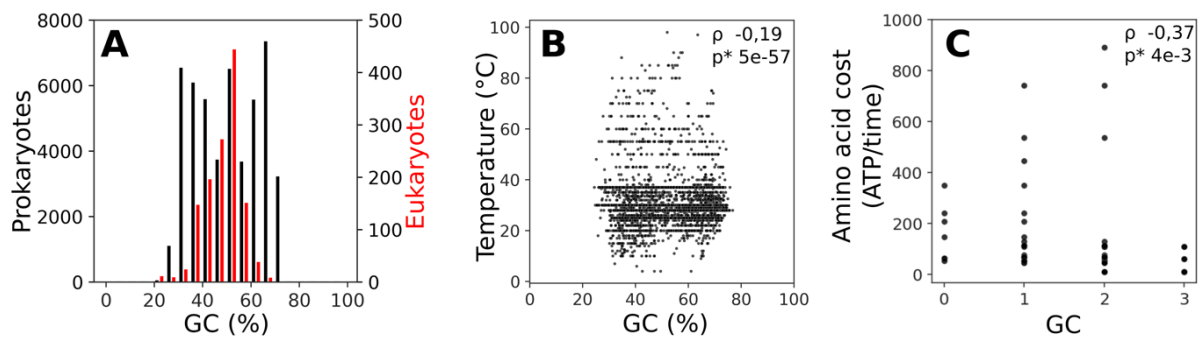

**Figure 1. GC content and associated traits.** (A) Histogram of GC content in 49783 prokaryotes (black bars) and 1322 eukaryotes (red bars). (B) Optimal growth temperature versus GC content in 6695 prokaryotes. (C) Amino acid metabolic cost versus GC content for the 61 protein-coding codons of the standard genetic code.

**GC content and reproductive traits in prokaryotes**

We have studied the relationship between the GC content in prokaryotic genomes and reproductive traits linked to r/K selection (Supplementary File 4). We expected r-strategists to produce cheaper, lower quality descendants at a faster rate (15, 16). This may take place in different ways because the cost of producing offspring depends on several factors, such as size of offspring, cost per unit volume and the amount of offspring produced per unit time. Descendants that are larger, more densely packed with energy and/or more abundant can be considered more expensive. We have examined these three factors in the case of prokaryotes. Figure 2 shows the relationship between the GC content and cell volume (panel A), average proteome cost per amino acid (panel B) and duplication time (panel C). Figure 2A shows a negative correlation between cell volume and GC content for 1278 prokaryotes (r -0.10, p* $6\cdot10^{-4}$). Figure 2B shows a strong negative correlation between average amino acid metabolic cost and GC content 49783 prokaryotes (r -0.88, p* $<10^{-300}$) when considering amino acid usage. Last, Figure 2C shows that duplication times of 21873 prokaryotes do not correlate with GC content (r $8.5\cdot10^{-5}$, p* 0.99). In sum, a high GC content is neutral to offspring cost in terms of the amount of offspring produced per unit time but correlates with cheaper descendants because of a lower average amino acid metabolic cost and, to some degree, to a smaller cell volume. Thus, the reproductive traits of prokaryotes link high GC content to the expectations for the ecological r strategy and lower GC content to the expectations for the ecological K strategy.

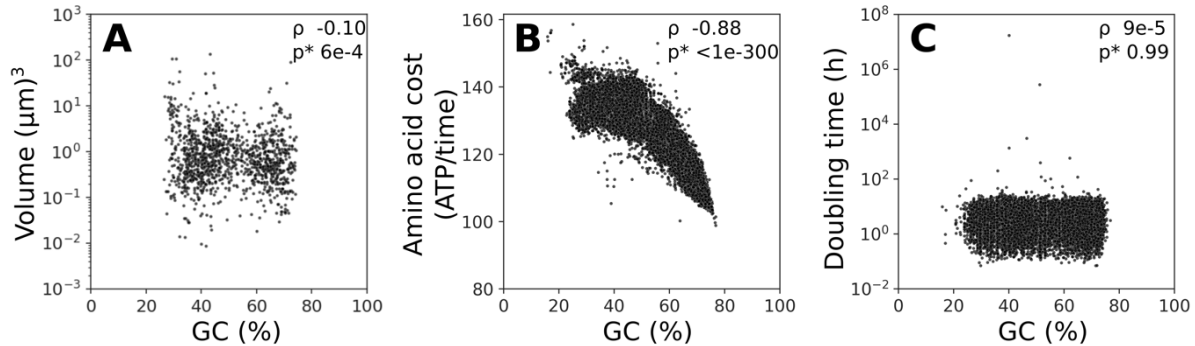

**Figure 2. Relationship between GC content and reproductive traits in prokaryotes.** (A) Cell volume versus GC content for 1278 prokaryotes. (B) Average amino acid metabolic cost versus GC content for 49783 prokaryotes. (C) Doubling time versus GC content for 21873 prokaryotes.

**GC content, cell shape and motility in prokaryotes**

The r strategy is often linked to colonization of new habitats through spatial mobility (17). This prompted us to look at the relationship between the GC content in prokaryotic genomes and traits linked to cell mobility (Supplementary File 4). In the case of prokaryotes, physics and hydrodynamic considerations decree that a rod-shaped bacillus form is better suited to detect nutrient gradients and for displacement relative to the spherical coccus form (42, 43). In addition, the energy cost of mobility is lessened by the presence of a flagellum (44). We analyzed phenotypic data on cell shape, motility and presence of a flagellum for more than three thousand prokaryotes. The data are shown in Figure 3 as notched boxplots for the different groups, where boxes and whiskers indicate the quartiles of each sample and a notch around the median provides a rough guide to statistical significance of the differences between group medians. Figure 3A shows that both the presence of motility and of a flagellum in prokaryotes correlate with higher GC content in a statistically significant manner in 3623 prokaryotes. Additionally, the GC content of prokaryotes with a bacillus form is higher than that of prokaryotes with a coccus form (Figure 3B) in 3445 prokaryotes. In all, a high GC content correlates with prokaryotes that can move in an energetically efficient manner. The cell shape and motility of prokaryotes links high GC content to colonization, which fits well with the ecological r strategy.

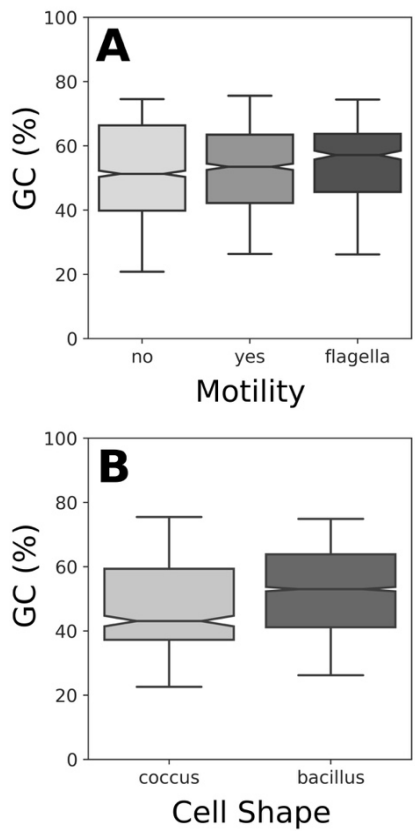

**Figure 3. Relationship between GC content, motility and cell shape in prokaryotes.** The notched box plots feature boxes and whiskers that indicate the quartiles of each sample and a notch around the median that provides a rough guide to statistical significance of the differences between medians. (A) Relationship between motility and GC content in 3623 prokaryotes. (B) Relationship between cell shape and GC content in 3445 prokaryotes.

**GC content and resource opportunism in prokaryotes**

Next, we examined the relationship between the GC content in prokaryotic genomes and metabolic traits linked to r/K selection (Supplementary File 4). r-strategists are often resource opportunists relative to K-strategists, i.e., they can use a higher number of different resources for growth (22, 29). In the case of prokaryotes, this is believed to come at the expense of a lower yield for nutrient to biomass conversion (17, 18). We gathered data for the number of different carbon substrates that can sustain growth in over 800 prokaryotes, directly evaluating metabolic capabilities (Figure 4A). Figure 4A shows that resource opportunism in prokaryotes correlates with higher GC content in a statistically significant manner (r 0.12, p* $2·10^{-7}$). Next, we evaluated the relationship between nutrient-to-biomass conversion and GC content. Carbon substrates with a heat of combustion lower than 9 kcal per gram of carbon lead to low biomass yields, while substrates with a heat of combustion over that threshold lead to high biomass yields (36). We calculated the median GC content for the prokaryotes that can grow on 66 different carbon substrates (Supplementary File 5) and plotted the relationship with the heat of combustion in Figure 4B. We observe that prokaryotes with a high GC content are associated to both low-yield and high-yield substrates, while prokaryotes with a low GC content are associated to high-yield substrates only. Panels A and B in Figure 4 are in line with r-strategists being resource opportunists, sometimes at the cost of a lower biomass yield, while K-strategists are specialized for the consumption of resources leading to a high biomass yield.

We reasoned that resource opportunism is the result of proteome phenotypic plasticity, which requires genes enabling metabolite transport and assimilation and a refined response to changes in the environment. We first gathered data for the number of coding sequences in a genome (Figure 4C), which has been related to general physiological plasticity (10–12). In the second place, we extracted the number of genes related to transport and metabolism of amino acids, nucleotides, carbohydrates, coenzymes, lipids, inorganic ions and secondary metabolites (Figure 4D). Since environment sensing mediates resource opportunism, in the third and fourth place we counted the number of genes related to signal transduction (Figure 4E) and the genes that mediate the resulting changes in protein expression (Supplementary Figure 1). All four genomic correlates of resource opportunism in prokaryotes correlate with higher GC content in a statistically significant manner (r ranges from 0.47 to 0.54, while the p*-values are $10^{-45}$ or lower). In sum, a high GC content correlates with prokaryotes with larger genomes that contain more genes dedicated to nutrient sensing and metabolism. Our data links high GC content in prokaryotes to a high degree of resource opportunism and a lower yield for nutrient to biomass conversion, which fits well with the ecological r strategy. A lower GC content is associated to a lower degree of resource opportunism and a higher nutrient-to-biomass yield, which fits the stable environments characteristic of the ecological K strategy.

We also included in the analysis the formation of endospores by some prokaryotes as a response to lack of nutrients in the environment (45). Endospores are resilient, dormant cells that require the presence of nutrients to reactivate and resume metabolic activity. In our view, the resource opportunism of r-strategists fits well with continued foraging in a changing environment, as shown in Figure 3. Conversely, we argue that endospore formation is better suited to K-strategists that can grow on a limited number of nutrients and lay dormant if these nutrients are absent to avoid the loss of energetically costly individuals. Figure 4F shows the relationship between endospore formation and GC content in 3072 prokaryotes. The data are shown as notched boxplots for the different groups as in Figure 3. Endospores are associated with a lower GC content in a statistically significant manner, which fits our expectations for the ecological K strategy.

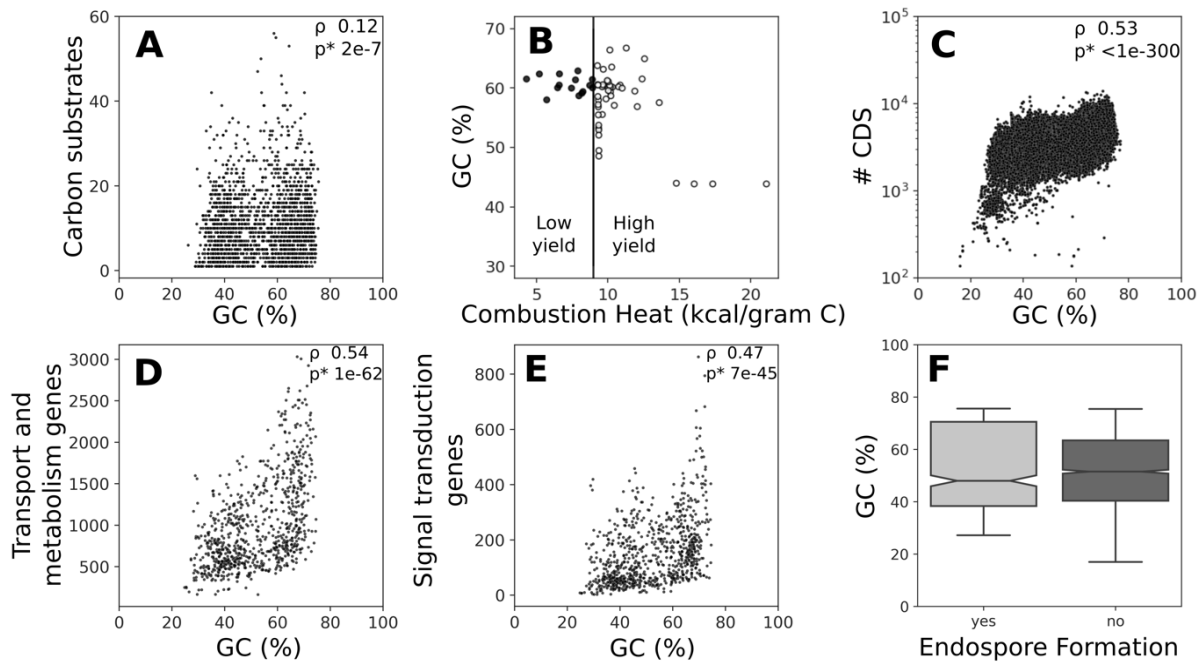

**Figure 4. Relationship between GC content and resource opportunism in prokaryotes.** (A) Number of different carbon substrates that can sustain growth versus GC content in 858 prokaryotes. (B) Relationship between the heat of combustion for 66 substrates able to sustain growth and the median GC content of the prokaryotes that can grow on each substrate. The number of prokaryotes used to calculate the median GC content for each substrate is at least 12. Substrates are classified in low and high yield according to (36) (pictured as black and white circles for clarity). (C) Number of coding sequences versus GC content in complete genomes of 46474 prokaryotes. (D) Number of genes related to transport and metabolism of amino acids, nucleotides, carbohydrates, coenzymes, lipids, inorganic ions and secondary metabolites versus GC content in 813 prokaryotes. (E) Number of genes related to signal transduction mechanisms versus GC content in 813 prokaryotes. (F) Relationship between endospore formation and GC content in 3072 prokaryotes.

**GC content and defense mechanisms in prokaryotes**

Last, we considered the relationship between the GC content in prokaryotic genomes and defense mechanisms (Supplementary File 4). Since r-strategists are able to thrive in unstable environments (30), it would seem reasonable that they harbor genes that can activate as a contextual response to a variety of biotic and abiotic hazards. We first evaluated the relationship between GC content and the number of genes in COG category V (defense mechanisms), which includes 175 clusters of orthologous genes. These gene clusters are related to the physiological response to biotic hazards such as bacteriophage infection (CRISPR, endonucleases and others), antibiotics and toxins (beta lactamases, drug pumps and others) and to abiotic hazards such as redox potential, desiccation, and toxic inorganic compounds. Figure 5A shows a positive correlation between GC content and the number of genes related to defense mechanisms for 811 prokaryotes (r 0.28, p* 7·$10^{-16}$). The correlations between GC content and the number of genes related to phenotypic plasticity (Figure 4D and 4E, Supplementary Figure 1, Figure 5A) could in principle be a consequence of a more general relationship between GC content and genome size (Figure 4C and (8)). However, not only the number of genes, but also the fraction of the total number of genes that is related to these functional categories increases with GC content (Supplementary Figure 2, r 0.43, p* 2·$10^{-36}$). We conclude that these correlations are not merely due to an increase in genome size but to GC content-dependent functional selection, as observed before for *Plasmodium* parasites (46). Our data links high GC content in prokaryotes to the ability to respond in a context-dependent manner to a higher number of environmental hazards, which fits the variable environments characteristic of the ecological r strategy.

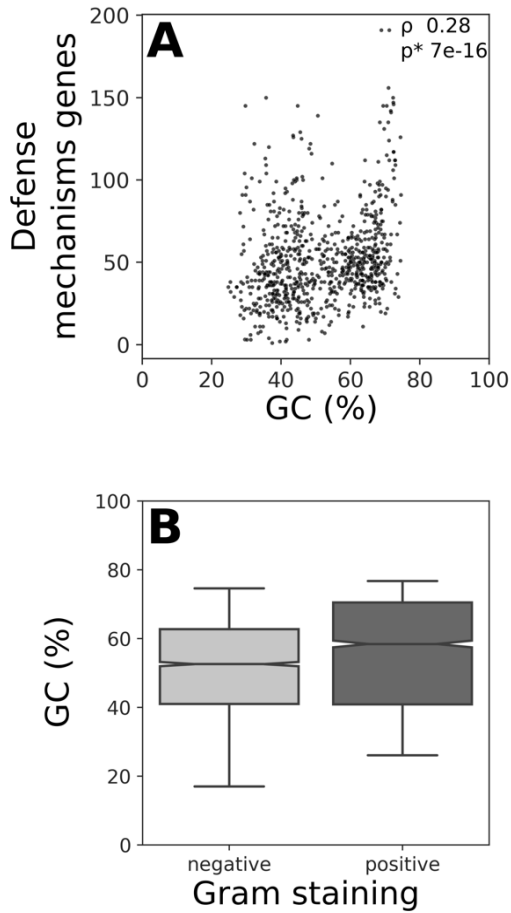

**Figure 5. Relationship between GC content and defense mechanisms in prokaryotes.** (A) Number of genes related to defense mechanisms versus GC content in 813 prokaryotes (B) Relationship between Gram staining and GC content in 5285 prokaryotes.

We have also specifically considered the outer cell membrane of Gram-negative prokaryotes as a large, permanent, and costly cellular structure relevant to defense and homeostasis maintenance. The outer cell membrane provides Gram-negative prokaryotes with an additional physical layer separating the inner membrane and the peptidoglycan cell wall from environmental hazards such as antibiotics (47) and other toxic compounds. Figure 5B shows the relationship between Gram staining and GC content in 5285 prokaryotes, with the data shown as notched boxplots as in Figure 3. Gram-negative bacteria are associated with a lower GC content than Gram-positive bacteria in a statistically significant manner. This is compatible with low GC content prokaryotic K-strategists producing offspring of higher quality and cost.

**Relationship with copiotrophs and oligotrophs**

Other ecological selection theories have been proposed to rationalize prokaryotic diversity. The copiotrophic/oligotrophic dichotomy was developed to understand the relationship between bacterial growth and nutrient availability (48). The basic tenet is that copiotrophs grow faster when nutrients are abundant, while oligotrophs grow faster when nutrients are scarce (49). Copiotrophs and r-strategists have multiple traits in common, such as larger genomes than oligotrophs and K-strategists (50), higher GC content (50), lower nutrient-to-biomass yield (18, 51), and a higher number of genes related to motility, transport and metabolism of nutrients, signal transduction, transcription and defense mechanisms (51, 52). Thus, many of the traits associated to a high GC content can be rationalized in terms of both r/K selection and the copiotrophy/oligotrophy axis. Copiotrophs and r-strategists are also expected to be optimized for fast growth (18), which should correlate with a higher number of ribosomal and transfer RNA genes (50, 51). The number of tRNA genes indeed shows a positive correlation with GC content (Supplementary Figure 3A, r 0.15, p* $5 \cdot 10^{-12}$), while the doubling times (Figure 2C) and the number of rRNA genes do not (Supplementary Figure 3B, r -0.077, p* $2 \cdot 10^{-3}$). Moreover, oligotrophs are expected to have a smaller cell size than copiotrophs due to a lower energy cost for maintenance (52, 53), which is at odds with the results in Figure 2A. These discrepancies between r/K selection and the copiotrophy/oligotrophy axis are not surprising, since they are built around two different main variables, namely environment stability and nutrient availability. In line with this, controlled laboratory experiments showed that these two selection theories are distinct in terms of functional gene variation (20). In sum, the traits associated to a high GC content can be partially but not completely rationalized in terms of the copiotrophy/oligotrophy axis.

## DISCUSSION

From the evidence in Figures 2 to 5, we propose that the GC content of prokaryotic genomes associates with multiple traits in r/K, generalist/specialist, plastic/fixist selection. GC content seems to have a stronger coupling with environment stability and its associated ecological variables than with nutrient availability and the copiotrophy/oligotrophy axis. We note that both GC content and r/K selection are continuous axes of variation and, consequently, individual organisms should not be classified as pure r- or K-strategists but rather placed along these axes at some distance of each archetype. We report statistically significant correlations of GC content with different traits relevant to r/K selection, with absolute values of the Spearman's rank correlation coefficient ranging from 0.10 to 0.88 (average 0.40±0.30) (Figures 2A, 2B, 4A, 4D, 4E and 5A). This indicates that the different traits associated with the r/K axis are subject to different degrees of selection and may help us reach a more nuanced definition of r/K selection in prokaryotes. Overall, we linked a high GC content in prokaryotic genomes to r-strategists with cheaper descendants thanks to a lower average amino acid metabolic cost and a smaller cell volume. These r-strategists are also colonizers of unstable environments with energetically efficient cell motility, thanks to flagella and a bacillus form. They are also generalists with a high degree of phenotypic plasticity that can grow on a larger number of carbon sources thanks to a larger number of genes dedicated to nutrient sensing and metabolism and are able to cope with various environmental hazards, thanks to a versatile repertoire of related genes. We note that the expression of many of these metabolism and defense-related genes and the activation of the flagellum is likely to take place only "on demand" in specific circumstances, which can keep at bay the energetic cost associated to the plasticity of r-strategists. Nevertheless, the advantages of the r-strategy seem to come at the expense of a lower nutrient-to-biomass yield. On the other hand, a low GC content in prokaryotic genomes is associated to K-strategists that produce high-cost offspring due to a higher average amino acid metabolic cost and a larger cell volume. These K-strategists are specialists for specific stable environments due to a lack of cell motility and a lower degree of phenotypic plasticity regarding resource assimilation and defense against environmental hazards. Specialization may enable the higher nutrient-to-biomass yield of these organisms. Low GC content prokaryotes maintain homeostasis via maintenance of a high-cost outer cell membrane and endospore formation as a response to nutrient deprivation. From a general perspective, we interpret that molecular diversity in the GC content of prokaryotes and the corresponding species diversity is partly due to ecological r/K selection in the context of a trade-off between growth in stable versus unstable environments (21). This could help explain the broader range of GC content in prokaryotes relative to eukaryotes (Figure 1).

Our results also bring together r/K selection with molecular traits such as the structure of the genetic code, genome size and GC content. A prokaryotic genome of any given GC content could in theory code for most of the traits associated here with the r and K strategies. However, the association between a high GC content and a cheaper offspring with a lower average amino acid metabolic cost (Figure 2B) is driven by the association between GC-rich codons and cheaper amino acids in the genetic code ((37) and Figure 1C) and therefore could not take place in the opposite direction. From this viewpoint, r/K selection on offspring cost can be regarded as one of the factors modulating GC content. Amino acid metabolism and the structure of the genetic code can be regarded as the molecular mechanisms dictating the direction of variation. In addition, laterally transferred DNA is more likely to lead to gene expression in genomes with a high GC content (54). This may drive the association between a high GC content and larger, more versatile genomes (Figure 4C and (8)). Thus, the sequence determinants of gene expression may also contribute to the alignment of genome size and r/K selection in prokaryotes.

We observed considerable heterogeneity in all traits studied here, at all GC contents (Figures 2 to 5). This indicates that GC content and r/K selection are not the sole determinants of diversity in prokaryotes. Future analysis of individual organisms, phylogenetic groups (55) and/or communities (56) may integrate GC content with additional traits, such as oxygen tolerance (12) and habitat preferences (57), in terms of additional selection theories such as the competitor-stress tolerator-ruderal framework (10, 55, 58), the yield-stress tolerance-resource acquisition framework (59), the resource acquisition-growth-maintenance framework (60), the competition-defense trade-off (61) and scramblers versus stayers (62). The comparison of GC content selection in prokaryotes versus eukaryotes (14, 56) may help us understand which selection trade-offs are present at each scale (10) and their interplay with phenotypic traits and the underlying molecular mechanisms.

## METHODS

### Genomic GC content

Reference (31) reports the genomic GC content for 2987 prokaryotes. We increased this number to 49783 prokaryotes in the following manner. We used the codon usage data in version o576245 of the CoCoPUTs 2 codon usage database (33), excluding from the analysis plasmids, organelles, viruses, organisms using non-standard genetic codes, eukaryotic genomes with less than 100 coding sequences reported, prokaryotic genomes with less than 1000 coding sequences reported, and any genome for which one or more codons were absent. If an organism is reported in both Genbank and Refseq, the Refseq data are used due to their better overall curation level. If a TaxID was associated to multiple genomic datasets in CoCoPUTs, as in multiple strains of the same species, we kept only the dataset with the highest number of coding sequences. These criteria allowed us to calculate the GC content of the coding sequences of 1322 eukaryotes, 49010 bacteria and 773 archaea. For the prokaryotes common to both datasets, coding sequence GC content closely follows genomic GC content (Supplementary Figure 4, r 0.99, p* <9·$10^{-300}$).

### Phenotypic traits

Data for Gram staining, sporulation, cell shape and motility, the presence/absence of a flagellum and the number of rRNA and tRNA genes were taken from (31). Cell volume data were calculated from the cell dimensions reported in (31), assuming an ellipsoidal shape. The number of different carbon substrates that can sustain growth for each prokaryote was calculated using the reports in (31). The heat of combustion for the different carbon substrates was calculated from the data in (35, 36).

The average metabolic cost of an amino acid was calculated by translating the codon frequencies in CoCoPUTs to amino acid frequencies using the standard genetic code. We then weighted the amino acid abundances by previously reported amino acid metabolic costs (32) to calculate the average metabolic cost of an amino acid in the corresponding proteome.

Reference (31) reports the growth temperature for 5671 prokaryotes. We increased this number to 6695 prokaryotes following a similar approach to (63). We manually curated the optimal growth temperature from BacDive (64), DSMZ (65), Pasteur Institute (PI), the National Institute for Environmental Studies (NIES) (66), and a curated list from a previous work (67). BacDive data are available through their API, which contains calls to retrieve the species list and to get all data about a specific species. We used we used previously published data files for DSMZ and PI (68) and scripts to query optimal growth temperature information in NIES (accessed July 2020). The agreement of the data in (31) and our newly compiled data for the overlapping organisms is statistically significant (Supplementary Figure 5A, r 0.91, p* <$10^{-300}$).

Reference (31) reports the doubling time for 408 prokaryotes. We increased this number to 21837 prokaryotes by retrieving the doubling times from EGGO DB (69) (accessed July 2021), which contains estimations of doubling time for genome sequences in RefSeq. We then mapped to current species names using GTDB (70) and retrieved the TaxIDs from the NCBI taxonomy. When there where multiple entries for a single species, we used the average value for duplication time. The agreement of the data in (31) and our newly compiled data for the overlapping organisms is statistically significant (Supplementary Figure 5B, r 0.25, p* 3·$10^{-7}$).

The number of coding sequences was taken from the CoCoPUTs database (33), excluding genomes for which the completeness level was not reported in the NCBI Assembly database (thus labelled as NN) (71).

The number of genes related to different functional categories (34) was taken from (12).

**Statistics**

We generated notched boxplots using the boxplot.stats function in R. The boxes and whiskers in the plot indicate the quartiles of each sample. The notch around the median extends to +/- 1.58 IQR/sqrt(n) of the median, with n being the number of organisms in a sample. It provides a rough guide to statistical significance of the differences between medians.

We calculated the p-values by testing if the correlation coefficients equal cero (Null Hypothesis). Coefficients were considered to have an approximate normal distribution under the null hypothesis, given the high number of organisms, and a variance of 1/(1-n). We used the Benjamini-Hochberg correction (72) to calculate p*-values that account for multiple testing.

**ACKNOWLEDGEMENTS**